\documentclass[showpacs,amsmath,amsart,twocolumn,showkeys,pra,superscriptaddress]{revtex4-1}
\usepackage{dcolumn}    
\usepackage{ifpdf}
\usepackage{amssymb,lineno,amsfonts}
\usepackage{graphicx}   
\usepackage{dcolumn}    
\usepackage{bm}         
\usepackage{bbm}
\usepackage{mathrsfs}
\usepackage{upgreek}
\usepackage{mathtools}
\usepackage{epstopdf}
\usepackage{setspace}
\usepackage{hyperref}
\usepackage{float}
\usepackage{natbib}
\usepackage[usenames,dvipsnames]{xcolor}
\usepackage[matrix,frame,arrow]{xypic}

\newcommand{\ket}[1]{\vert{#1}\rangle}

\newcommand{\proj}[1]{\vert{#1}\rangle\langle{#1}\vert}

\definecolor{med-blue}{RGB}{25,25,112}
\hypersetup{colorlinks, linkcolor={red},citecolor={blue}, urlcolor={MidnightBlue}}

\begin{document}

\title{Discriminating between L\"uders and von Neumann  measuring devices: \\ An NMR investigation }
\author{C. S. Sudheer Kumar, Abhishek Shukla, and T. S. Mahesh
}
\email{mahesh.ts@iiserpune.ac.in}
\affiliation{Department of Physics and NMR Research Center,\\
	Indian Institute of Science Education and Research, Pune 411008, India}

\begin{abstract}
	Measurement of an observable on a quantum system involves a probabilistic  collapse of the quantum state and a corresponding measurement outcome.  L\"uders and von Neumann state update rules attempt to describe the above phenomenological observations.  These rules are identical for a nondegenerate observable, but differ for a degenerate observable. While L\"uders rule preserves superpositions within a degenerate subspace under a measurement of the corresponding degenerate observable, the von Neumann rule does not.  Recently Hegerfeldt and Mayato 
	[Phys. Rev. A, 85, 032116 (2012)]
	had formulated a protocol to discriminate between the two types of measuring devices.  Here we have reformulated this protocol for quantum registers comprising of system and ancilla qubits.  We then experimentally investigated this protocol using nulear spin systems with the help of NMR techniques, and found that  L\"uders rule is favoured.
\end{abstract}

\maketitle
\section{Introduction}
Quantum measurement paradox lies at the heart of foundations of quantum mechanics\cite{D_Home_book}. It's an experimental fact that, upon measurement, a quantum state collapses into an eigenstate of the observable being measured. However there is no collapse in the unitary evolution described by Schr\"odinger equation, and therefore, the collapse has to be imposed from outside the formalism.  

\begin{figure}
	\hspace*{-0.3cm}
	\centering
	\includegraphics[width=9cm,clip = true,trim = 1.2cm 6.8cm 1.8cm 2cm]{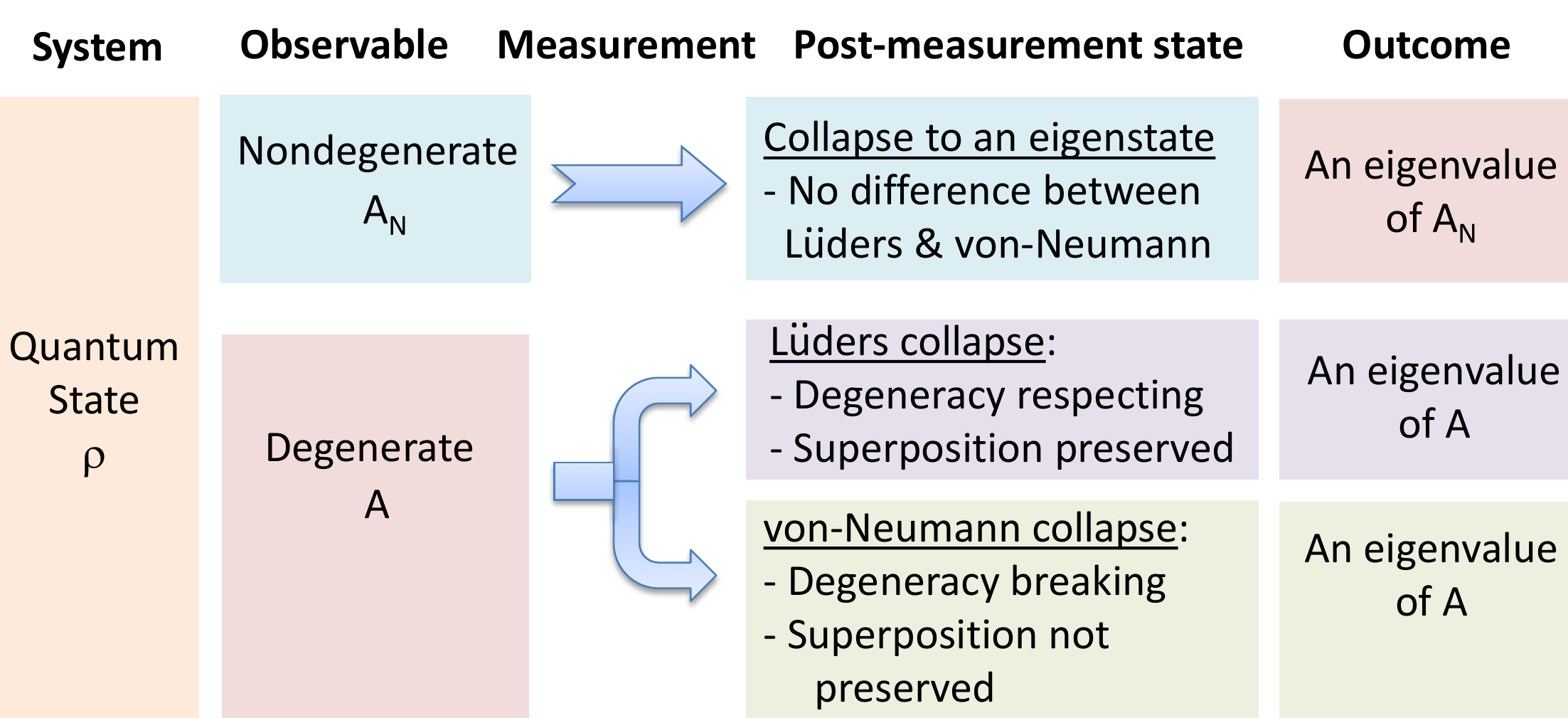}
	\caption{Comparison between L\"uders and von Neumann measurement postulates. }
	\label{luder}
\end{figure}

Let us assume an observable $A_N$ with discrete and nondegenerate eigenspectrum.  In that case, the measurement leads to a collapse of the state to one of the eigenstates of $A_N$ (see Fig. \ref{luder}).
On the other hand, if we consider an observable $A$ with a degenerate eigenspectrum, there are two extreme rules to update the state after the measurement. The most commonly used rule was postulated by Gerhart L\"uders in 1951 \cite{luders1951zustandsanderung,Luders_originl_paper}.  According to it, a system existing in a superposition of degenerate eigenstates is unaffected by the measurement such that the superposition is preserved. 
However, an earlier postulate by von Neumann, proposed in 1932  \cite{von_math_foundof_QM}, does not preserve such a superposition. In the latter postulate,
the measuring device refines the observable $A$ into another commuting observable $A'$ (actual system observable) having a nondegenerate spectrum.  The resulting measurement collapses the state to an eigenstate of $A'$, and the original superposition is not preserved under the measurement as if the degeneracy has been lifted \cite{Discriminate_von_Luder_protocol}. 

Although, one generally assumes L\"uders state update rule implicitly in quantum physics, occassionally one encounters applications of the von Neumann state update rule.
One example is in the context of Leggett-Garg inequality in multilevel quantum systems \cite{multi_level_LGI}.
In principle, measurements which are intermediate between L\"uders and von Neumann can also be conceived \cite{Discriminate_von_Luder_protocol,multi_level_LGI}. 

\begin{figure}
	\centering
	\includegraphics[width=9cm,clip = true,trim = 0.6cm 0.8cm 2.5cm 0.5cm]{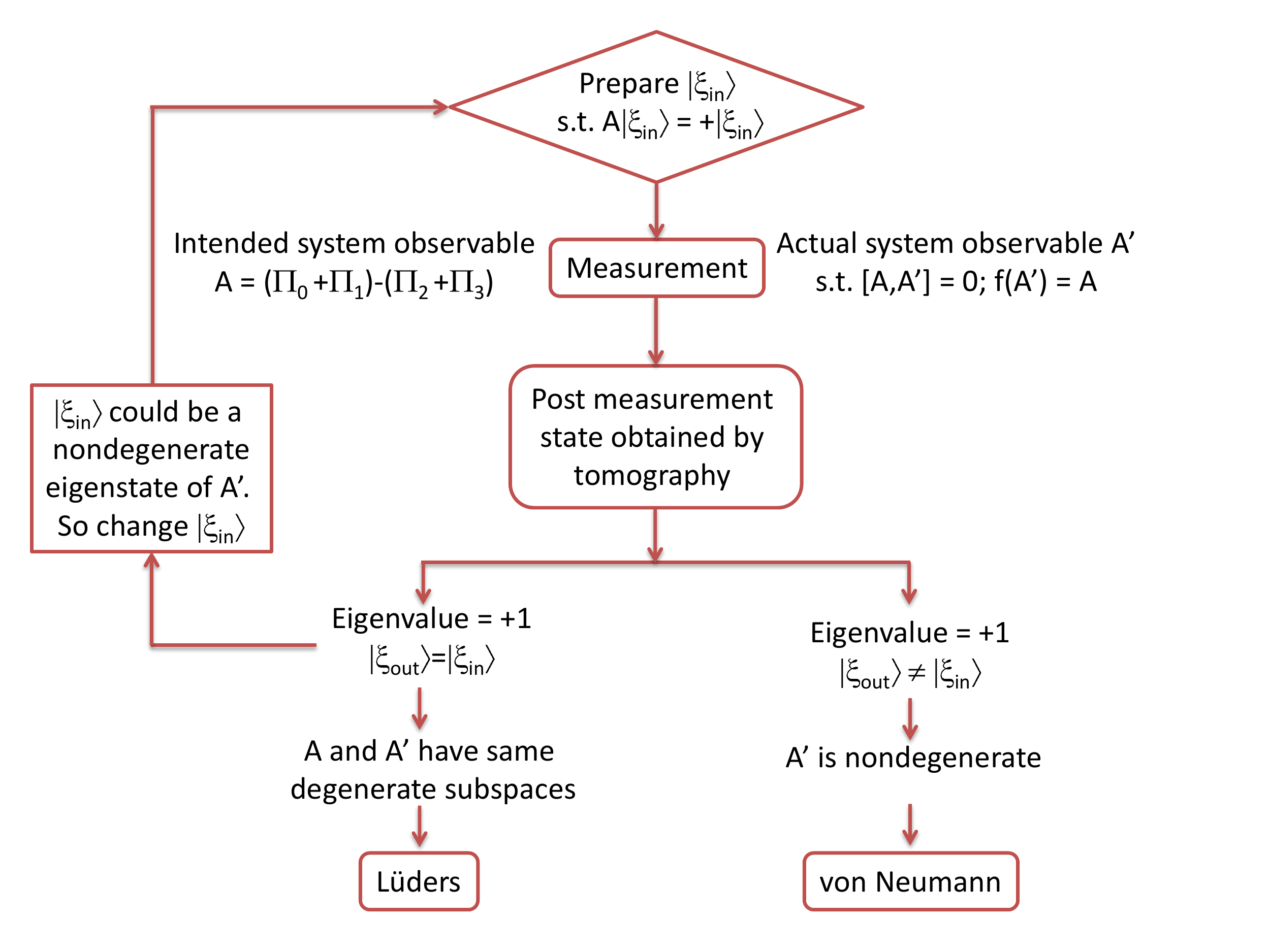}
	\caption{HM protocol for discriminating between L\"uders and von Neumann measurements. }
	\label{HMprotocolfig}
\end{figure}

Recently, Hegerfeldt and Mayato  have proposed a general protocol (HM protocol) to discriminate between L\"uders and von Neumann kind of measuring devices \cite{Discriminate_von_Luder_protocol}. 
To explain this protocol we consider an observable $A$, having two-fold degenerate eigenvalues, say $+1$ and $-1$ (see Fig. \ref{HMprotocolfig}). The HM protocol involves the following steps: (i) prepare an eigenstate $\ket{\xi_\mathrm{in}}$ of $A$, (ii) let the device measure $A$, and (iii) characterize the output state.  In step (ii) a L\"uders measurement will preserve the state, while a von Neumann measurement may not.  The last step is simply to determine if the step (ii) has changed the state or not.  If the state has changed, we conclude that the device is von Neumann.  Else, either the device is of L\"uders type, or the chosen initial state $\ket{\xi_\mathrm{in}}$ happens to be a nondegenerate eigenstate of the actual system observable $A'$.  To rule out the latter possibility, one may change the initial state and repeat the above steps (Fig. \ref{HMprotocolfig}).  This way one can attempt to discriminate between the L\"uders and von Neumann measurement devices.

In this work, we reformulate the HM protocol for a quantum register and try to investigate it using experiments.  Nuclear spin ensembles in liquid, liquid-crystalline, or solid-state systems have often been chosen as convenient testbeds for studying foundations of quantum physics \cite{suter1988study,Moussa,LGI_Soumya,Context_cssk}.  Their main advantages are long coherence times and excellent control over quantum dynamics via highly developed nuclear magnetic resonance (NMR) techniques.

In section II, we briefly explain the HM protocol as adapted to an NMR setup. 
The experimental details to discriminate between the L\"uders and von Neumann   measuring devices is described in section III.  Finally we conclude in section IV.

\section{Theory}
For the sake of clarity, and also to match the experimental details described in the next section, we consider a system of two qubits.
 Since the system is to be measured projectively, dimension of the pointer basis should be greater than or equal to that of the system, and hence we need at least two ancillary qubits. 
We refer to the ancillary qubits as (1,2) and  system qubits as (3,4). 
We use Zeeman product basis as our computational basis and denote eigenkets of $\sigma_z$, the Pauli $z$-operator,
by $\ket{0}$ and $\ket{1}$. We denote the basis vectors of system qubits as 
\begin{eqnarray}
\ket{\phi_{0}} = \ket{00}, \ket{\phi_{1}} = \ket{01}, \ket{\phi_{2}} = \ket{10}, 
\ket{\phi_{3}} = \ket{11}.
\label{phis}
\end{eqnarray}

Let us assume a two-fold degenerate system-observable with spectral decomposition
\begin{eqnarray}
A  = (\Pi_0+\Pi_1)-(\Pi_2+\Pi_3), ~~~~
\label{A}
\end{eqnarray}
where the projectors are defined as $\Pi_j=\proj{\chi_j},\ket{\chi_0}=\alpha_0\ket{\phi_0}+\beta_0\ket{\phi_1}$, $\ket{\chi_1}=\alpha_1\ket{\phi_0}+\beta_1\ket{\phi_1}$, $\ket{\chi_2}=\alpha_2\ket{\phi_2}+\beta_2\ket{\phi_3}$, $\ket{\chi_3}=\alpha_3\ket{\phi_2}+\beta_3\ket{\phi_3}$ are eigenvectors of $A$.
 The projectors have the property $\Pi_k\Pi_l=\delta_{kl}\Pi_k$. We note that $A$ has no unique spectral decomposition due to the degeneracy.

We consider a measurement model, wherein a quantum system being measured undergoes a joint evolution with the measuring device, ultimately forming an entangled state.  When the measuring device collapses to a particular pointer state, the system also collapses to the corresponding eigenstate.  Let $Q$ be the observable corresponding to the ancilla (measuring device) and $g$ be the system-ancilla interaction strength.  The joint evolution is then of the form
\begin{eqnarray}
U_\mathrm{int} = \exp(-i{\cal H}_\mathrm{int}\tau),
\end{eqnarray}
where ${\cal H}_\mathrm{int} = g ~Q \otimes A$ is the interaction Hamiltonian in units of angular frequency.

To fix the basis inside a degenerate subspace, we should choose a nondegenerate observable $A'$ which commutes with $A$, so that they are simultaneously diagonalizable and hence we can find a common eigenbasis. For simplicity we choose the computional basis $\{\ket{\phi_j}\}$ as the common eigenbasis. Then the observable $A'$ must have the following spectral decomposition
\begin{eqnarray}
A'= \sum_{j=0}^3 a'_j P_j,
\label{A'_defined}
\end{eqnarray} 
where $P_j = \proj{\phi_j}$ and the nondegenerate eigenvalues $a'_j$ are yet to be determined. 

Let us assume the device to be von Neumann which refines the degenerate observable $A$ that is being measured, into a nondegenerate observable $A'$, via a mapping $f(A')=A$. 
As the refined observable $A'$ has nondegenerate eigenvalues and commutes with $A$, it fixes the basis inside the degenerate subspace. However, the choice of $A'$ is not unique, i.e., any orthonormal basis inside the degenerate subspace can be nondegenerate eigenkets of $A'$, and the von Neumann device has the freedom to  choose among them \cite{von_math_foundof_QM}.

The measurement outcome is passed via the refining function $f$, such that $f(a'_0)=f(a'_1)=+1$ and $f(a'_2)=f(a'_3)=-1$. Hence the outcome is same as if $A$ is being measured. To projectively measure the observable $A'$,   the measuring device has to jointly evolve with the system under the interaction Hamiltonian, 
\begin{eqnarray}
{\cal H}'_\mathrm{int}= g ~Q\otimes A'.
\end{eqnarray}
For instance, we choose $Q=q_1\sigma_{1z}+q_2\sigma_{2z}$, where 
$\sigma_{1z} = \sigma_z \otimes\mathbbm{1}_2$, 
$\sigma_{2z} = \mathbbm{1}_2\otimes\sigma_z$ and $\mathbbm{1}_2$ is $2\times 2$ identity operator.
The joint evolution between the measuring device (ancillary qubits) and the system is described by the unitary operator
\begin{eqnarray}
U'_\mathrm{int} = \exp(-i\cal{H}_\mathrm{int}' \tau),
\end{eqnarray}
where $\tau$ is duration of the evolution.

\begin{figure}[b]
	\centering
	\includegraphics[width=7cm,clip = true,trim = 5cm 5.5cm 5cm 6cm]{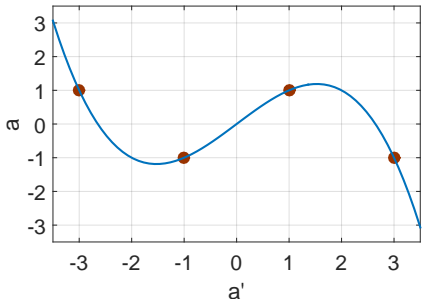}
	\caption{An interpolating function $a = f(a') = (-a'^3+7a')/6$ mapping the nondegenerate eigenvalues $a'$ of $A'$ onto degenerate eigenvalues $a$ of $A$.}
	\label{fofx}
\end{figure}

If each of the quantum register is initially prepared in $\ket{\Phi_0} = \ket{++++}$, with $\ket{+} = (\ket{0}+\ket{1})/\sqrt{2}$, the state after the joint evolution is given by
\begin{eqnarray}
U'_\mathrm{int}\ket{\Phi_0}   &=&  \frac{1}{2} \bigg( e^{-iga'_0Q \tau}\ket{++}\ket{\phi_0}+   
  e^{-iga'_1Q \tau}\ket{++}\ket{\phi_1}+ \nonumber \\ 
&&   e^{-iga'_2Q \tau}\ket{++}\ket{\phi_2}+   
   e^{-iga'_3Q \tau}\ket{++}\ket{\phi_3} \bigg) \nonumber \\
&=&  \frac{1}{2} \sum_{j=0}^3 \ket{\psi_j}\ket{\phi_j},
\label{von_evolutn}
\end{eqnarray}
where $\ket{\phi_j}$ are as defined in Eqs. \ref{phis} and $\ket{\psi_j} = \exp(-iga'_jQ \tau)\ket{++}$ represent states of the ancillary qubits.  To realize the projective measurement, the pointer basis $\{\ket{\psi_j}\}$ must be orthonormal.  Imposing the mutual orthogonality condition results in trigonometric constraint equations leading to a set of possible solutions.  One such possible solution is
\begin{eqnarray}
\begin{array}{l|l}
a'_0 = -a'_2 = -3 & q_1 = \pi/(4g\tau) \\
a'_1 = -a'_3 = 1 & q_2 = -q_1/2.\\
\end{array}
\end{eqnarray}
  Again, the von Neumann measuring device has the freedom to choose a particular pointer basis among several possible ones.   Substituting the above values in Eq. \ref{A'_defined}, we obtain,
  \begin{eqnarray}
A'=-3 P_0+ P_1 +3 P_2 -P_3,
  \end{eqnarray}
  which is obviously nondegenerate in the computational basis.  The refining function $f$ can now be setup by interpolating the eigenvalue distribution (see Fig. \ref{fofx}).  For the above example, we find a possible map to be $f(A') = (-A'^3+7A')/6 = A$.

\begin{figure}
  	\centering
  	\includegraphics[width=9cm,clip = true,trim = 1cm 0cm 0cm 0cm]{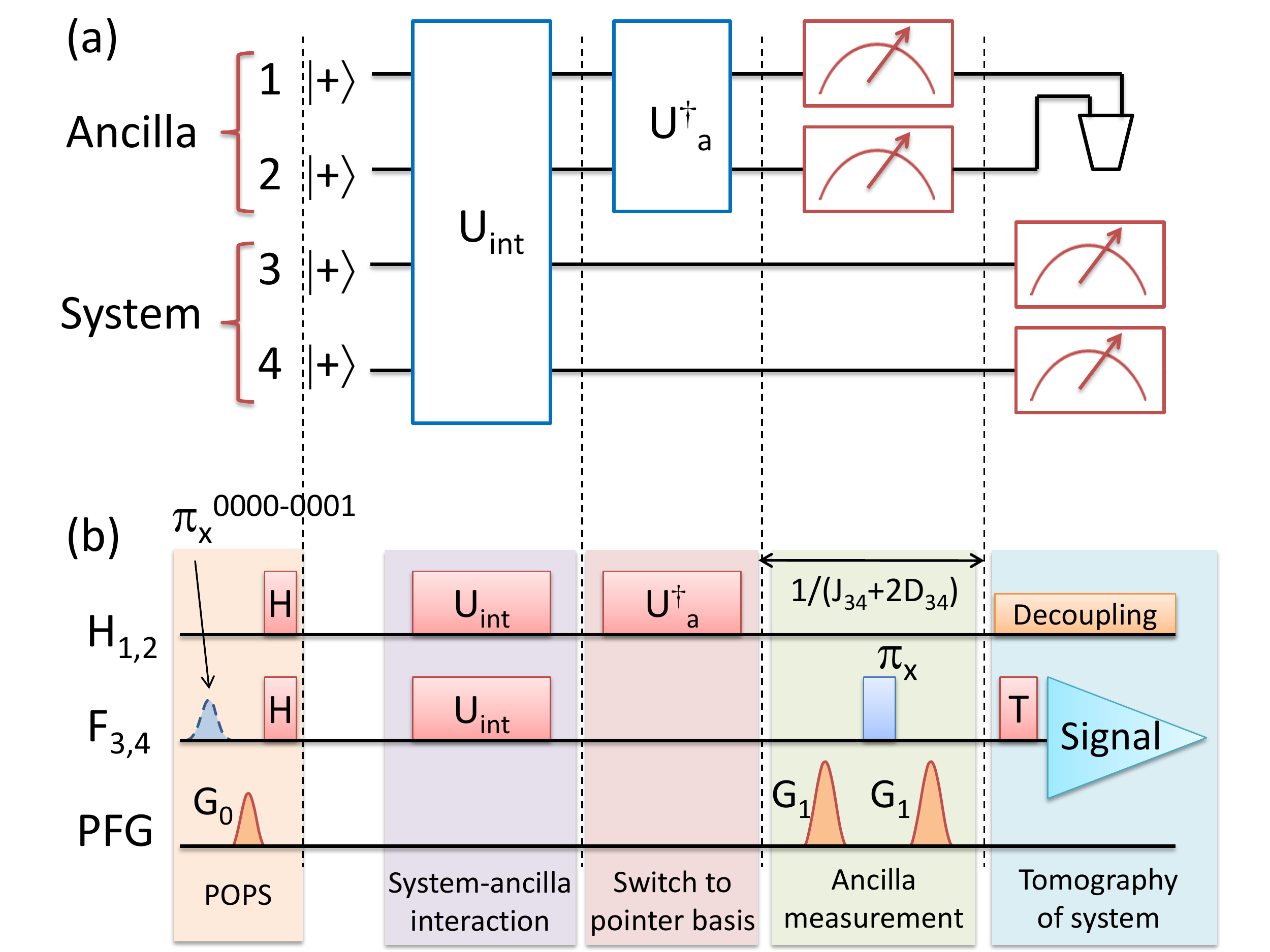}
  	\caption{(a) Quantum circuit to discriminate L\"uders and von Neumann devices. 
  		(b) The NMR pulse-scheme to implement the circuit in (a).}
  	\label{fig_discrimnt_von_lud}
  \end{figure}

   The quantum circuit for discriminating L\"uders and von Neumann devices, illustrated in Fig. \ref{fig_discrimnt_von_lud}, involves four qubits each of which is initialized in state $\ket{+}$.  If the device is L\"uders, the system undergoes a joint evolution $U_\mathrm{int}$ with the ancilla resulting in the state
   \begin{eqnarray}
 U_\mathrm{int} \ket{\Phi_0} &=& \frac{1}{\sqrt{2}}
 \bigg( e^{-igQ \tau}\ket{++} \frac{d_0\ket{\chi_0}+d_1\ket{\chi_1}}{\sqrt{2}} + \nonumber \\
 && e^{igQ \tau}\ket{++} \frac{d_2\ket{\chi_2}+d_3\ket{\chi_3}}{\sqrt{2}} \bigg) \nonumber \\
  & = & \frac{1}{\sqrt{2}} \bigg( 
  \ket{\psi_1} \frac{\ket{\phi_0}+\ket{\phi_1}}{\sqrt{2}} +
 \ket{\psi_3} \frac{\ket{\phi_2}+\ket{\phi_3}}{\sqrt{2}}
 \bigg),~~~~
\label{Luder_evolution}
\end{eqnarray}
 where the coefficients $d_j$ depend on the choice of $\ket{\chi_j}$ (defined after Eq. \ref{A}) \cite{ludernote1}.

 Note that if the measuring device is of von Neumann type, it will instead measure $A'$, and pass the measurement outcome via the function $f$, as explained before.  
 
After the joint evolution of system and ancilla, a selective measurement of ancilla qubits is carried out.
Generally in a quantum measurement the measuring device collapses to its pointer basis.  In our scheme, we perform the projective measurement in the computational basis after transforming the ancilla qubits onto the computational basis using a similarity transformation $U_{a}^\dagger$, such that
\begin{eqnarray}
U_{a}\ket{00}=\ket{\psi_0},U_{a}\ket{01}=\ket{\psi_1},\nonumber\\
U_{a}\ket{10}=\ket{\psi_2},U_{a}\ket{11}=\ket{\psi_3}.
\label{U_ra_constraints}
\end{eqnarray}
By substituting the explicit forms of $\ket{\psi_j}$, we obtain
\begin{eqnarray}
U_{a}=\frac{1}{2}
\left[
\begin{array}{cccc}
z^3 & z^{-1} & z^{-3} & z \\
z^{9} & z^{-3} & z^{-9} & z^{3} \\
z^{-9} & z^{3} & z^{9} & z^{-3} \\
z^{-3} & z & z^{3} & z^{-1}
\end{array}
\right],
\label{U_ra}
\end{eqnarray}
where $z=\exp(i\pi/8)$. 

Finally, the ancilla is traced-out and the state of system qubits is characterized with the help of quantum state tomography.

According to the L\"uders state update rule, if a degenerate observable $A$ (as in Eq. \ref{A}) is measured on a system in state $\rho_0$, then the postmeasurement state of the ensemble is described by
\begin{eqnarray}
\rho_{L}=\sum_{l=\pm 1}{\mathbb P}_l\rho_{0}{\mathbb P}_l,
\label{rhoLgen}
\end{eqnarray}
 where ${\mathbb P}_{+1} = \Pi_{0}+\Pi_{1}$,
${\mathbb P}_{+1} = \Pi_{2}+\Pi_{3}$.
For the initial state $\rho_0 = \proj{\Phi_0}$,
we obtain 
\begin{eqnarray}
\rho_L = 
(\mathbbm{1}_4 + \mathbbm{1}_2 \otimes \sigma_x)/4.
\label{rhoL}
\end{eqnarray}

However according to von Neumann's degeneracy breaking state update rule, the postmeasurement state of the ensemble is given by
\begin{eqnarray}
\rho_N = \sum_{j=0}^3 \Pi_j \rho_0 \Pi_j, 
\label{rhoNgen}
\end{eqnarray}
where, $\Pi_j$'s are fixed by the refining observable $A'$.
Therefore, for the initial state  $\rho_0 = \proj{\Phi_0}$ and the observable $A'$ (Eq. \ref{A'_defined}), the postmeasurement state collapses to a maximally mixed state, i.e.,
\begin{eqnarray}
\rho_N = \mathbbm{1}_4/4.
\label{rhoN}
\end{eqnarray}
 In both the cases, the probabilities of obtaining the eigenvalues $\pm 1$ are identical, i.e.,
\begin{eqnarray}
p_{+1} &=& \mathrm{Tr}({\mathbb P}_{+1}\rho_{0}{\mathbb P}_{+1})=\sum_{j=0,1}\mathrm{Tr}(\Pi_j \rho_0 \Pi_j) ~~\mathrm{and,}\nonumber \\ 
p_{-1} &=& \mathrm{Tr}({\mathbb P}_{-1}\rho_{0}{\mathbb P}_{-1})=\sum_{j=2,3}\mathrm{Tr}(\Pi_j \rho_0\Pi_j).
\end{eqnarray}

Thus although, the measurement outcomes (eigenvalues) and their probabilities are identical, the postmeasurement states $\rho_L$ and $\rho_N$ are different \cite{Luders_originl_paper,von_math_foundof_QM,Discriminate_von_Luder_protocol,multi_level_LGI}.
In fact, the Uhlmann fidelity between $\rho_L$ and $\rho_N$ turns out to be $F(\rho_L,\rho_N)=\mbox{Tr}\sqrt{\sqrt{\rho_N}\rho_L\sqrt{\rho_N}}=1/\sqrt{2}$ \cite{quant_info_neilson_chuang}.
Therefore, it is possible to discriminate between the Luders and von Neumann devices by simply characterizing the final state of the system as shown by the circuit in Fig. \ref{fig_discrimnt_von_lud}.

\begin{figure}
	\centering
	\includegraphics[width=7.5cm,clip = true,trim = 5cm 5cm 3.5cm 3.5cm]{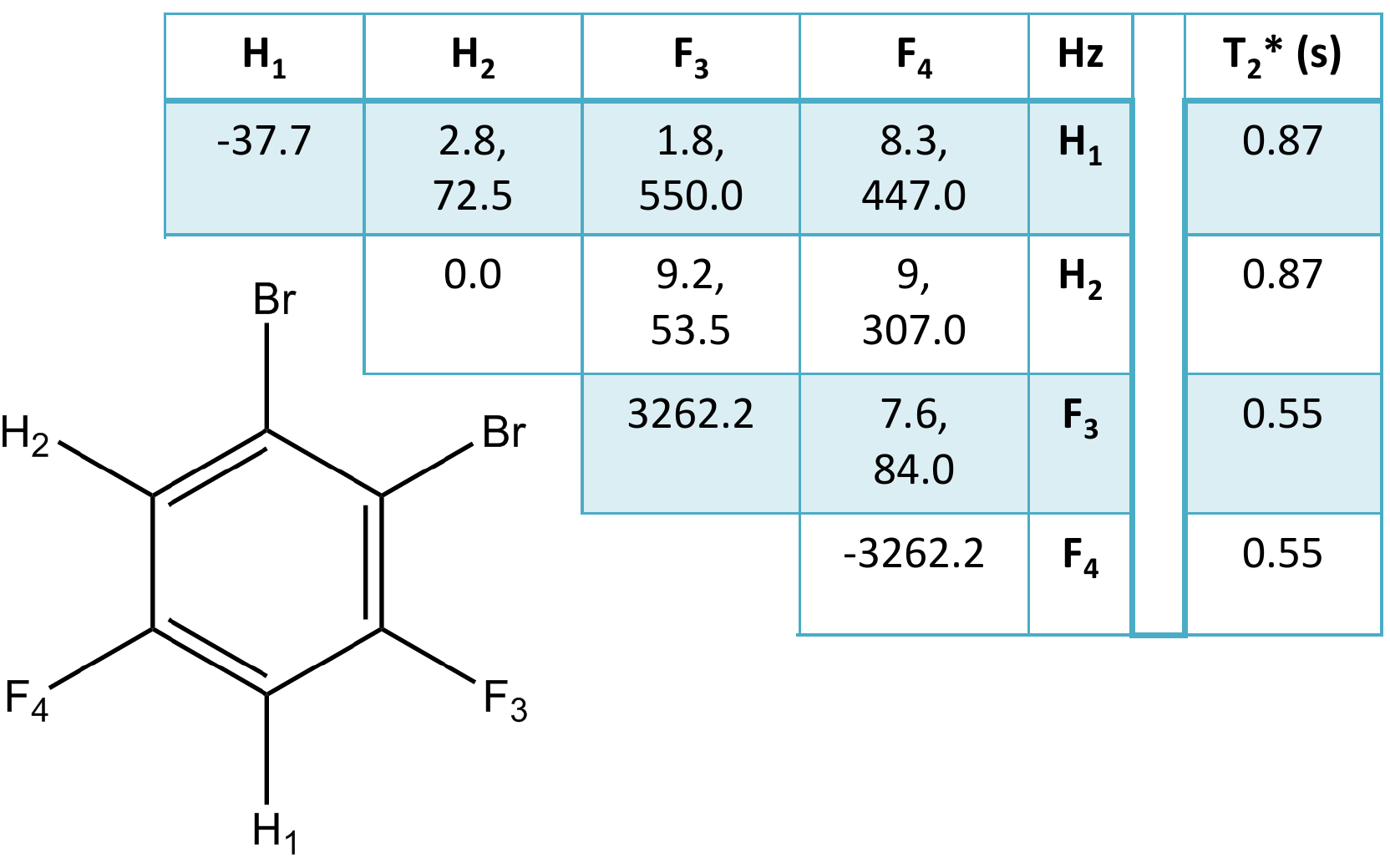}
	\caption{Molecular structure of 1,2-Dibromo-3,5-difluorobenzene, Hamiltonian parameters, and the relaxation parameters.  In the table, the diagonal values indicate resonance offsets ($\omega_j/2\pi$); off-diagonal values $(J_{ij},D_{ij})$ indicate the indirect and the residual direct spin-spin coupling constants respectively (in Hz); the last column lists approximate effective transverse relaxation time constants ($T_2^*$).}
	\label{mol_H}
\end{figure}

\section{Experiment}
 We utilize the four spin-1/2 nuclei of 1,2-dibromo-3,5-difluorobenzene (DBDF) as our quantum register.  
About 12 mg of DBDF was partially oriented in 600 $\upmu$l of liquid crystal MBBA.  The molecular structure of DBDF and its NMR Hamiltonian parameters  are shown in Fig. \ref{mol_H}.  
The experiments were performed at 300 K on a 500 MHz Bruker UltraShield NMR spectrometer.

The secular part of the spin-Hamiltonian is of the form  \cite{cavanagh},
\begin{eqnarray}
{\cal H}_0 &=& -\sum_{j=1}^4 \omega_j I_{jz} 
+ 2\pi\sum_{j,k>j} (J_{jk}+2D_{jk}) I_{jz}I_{kz} \nonumber \\
&&+ 2\pi (J_{12}-D_{12})( I_{1x}I_{2x}+ I_{1y}I_{2y}),
\end{eqnarray}
where $\omega_j$, $J_{ij}$, and $D_{ij}$ are the resonance off-sets,  indirect scalar coupling constants, and direct dipole-dipole coupling constants (Fig. \ref{mol_H}). 
The strong-coupling term (i.e., the last term) is relevant only for (H$_1$, H$_2$) spins since $\vert\omega_1-\omega_2\vert < 2\pi \vert D_{12} \vert$.
We choose H$_1$, H$_2$ as ancilla (qubits 1, 2) and F$_3$, F$_4$ as the system (qubits 3, 4).

The NMR pulse diagram to implement the quantum circuit in Fig. \ref{fig_discrimnt_von_lud}(a) is shown
in Fig. \ref{fig_discrimnt_von_lud}(b).  It begins with the initial state preparation. The thermal equilibrium state of the NMR system in the Zeeman eigenbasis under high-field, high-temperature, and secular approximation is given by \cite{Levitt_Spindynabook,cavanagh},
\begin{equation}
\rho_{\mathrm{eq}} = 
\mathbbm{1}_{16}/16 + \sum_{j=1}^4 \epsilon_j I_{zj},
\label{rho_eq}
\end{equation}
where $\epsilon_j \sim 10^{-5}$ are the purity factors and the second term in the right hand side corresponds to the traceless deviation density matrix.  The identity part is invariant under the unitary transformations and  does not give rise to observable signal. Therefore only the deviation part is generally considered for both state preparation and characterization \cite{cory}.

The initial state of the quantum register assumed in the theory section, i.e., $\ket{\Phi_0}$ can be prepared by applying an Hadamard operator on each of the four qubits in a pure $\ket{0}$ state. However, in NMR, the preparation of such pure states is difficult and instead a pseudopure state is used \cite{cory}.
In our work, we utilize a technique based on preparing a pair of pseudopure states (POPS) \cite{POPS_Fung}. It involves inverting a single transition  and subtracting the resulting spectrum from that of the thermal equilibrium.  By inverting the transition $\ket{0000}$ to $\ket{0001}$ transition using a transition selective $\pi$ pulse, followed by  Hadamard gates ({\textsf H}) on all the spins we obtain the POPS deviation density matrix: 
\begin{eqnarray}
\rho_\mathrm{POPS} =  \big(\proj{++++}-\proj{+++-}\big) \nonumber.
\end{eqnarray}

We then implemented the quantum circuit shown in Fig. \ref{fig_discrimnt_von_lud} (a) using the pulse sequence in Fig. \ref{fig_discrimnt_von_lud} (b).  
As evident from circuit in Fig. \ref{fig_discrimnt_von_lud},  controls are designed to implement $U_{\mathrm{int}}$ (Eq. \ref{Luder_evolution}) since we  intend to measure $A$. Whether to map it to $A'$ or not is left to the device.
The unitary operators $U_\mathrm{int}$ and $U_{a}^\dagger$ were realized by 
bang-bang optimal control \cite{Bangbang_TSM}. Hadamard and tomography operations were only few hundred micro seconds long and had a simulated fidelity of about 0.99, when averaged over $\pm 10\%$ inhomogeneous RF fields.
The combined operation of $U_\mathrm{int}$ and $U_a^\dagger$ was about 17 ms in duration and had an average fidelity over $0.933$.

The intermediate measurement on ancilla was realized by applying strong pulse-field-gradients (PFG). By applying a $\pi_x$ pulse on the system spins in between two symmetrically spaced PFG pulses, we realize the selective dephasing of the ancilla spins (Fig. \ref{fig_discrimnt_von_lud} (b)).  The central $\pi_x$ also refocuses all the system-ancilla coherent evolutions during the ancilla measurement. When averaged over the sample volume this process retains only the diagonal terms in the density matrix of the ancialla spins and thus simulates a projective measurement of ancilla.  Setting the total duration of this process to $1/(J_{34}+2D_{34})$ also ensures refocusing of (F$_3$, F$_4$) interactions.  

Finally, the density matrix of the system qubits was characterized using quantum state tomography.  It involved nine independent  measurements with different tomography pulses ({\textsf T}) (Fig. \ref{fig_discrimnt_von_lud} (b)) \cite{Tomography_Chuang447,Tomo_Soumya}.

The results of the quantum circuit (Fig. \ref{fig_discrimnt_von_lud}) on $\proj{++++}$ state by L\"uders and von Neumann devices are described in Eqs. \ref{rhoL} and \ref{rhoN} respectively.  
For L\"uders measurement with the POPS input state
$\proj{++++}-\proj{+++-}$, the final deviation density matrix (in circuit \ref{fig_discrimnt_von_lud}) is expected to be 
\begin{eqnarray}
\rho_L' = \mathbbm{1}_2 \otimes \sigma_x/2.
\label{Lud_finalstate}
\end{eqnarray}
On the other hand, for von Neumann measurement, the POPS input state leads to a maximally mixed final state with a null deviation density matrix $(\rho_N')$.

\begin{figure}
\centering
\includegraphics[width=7cm,clip = true,trim = 0cm 6cm 0cm 5.5cm]{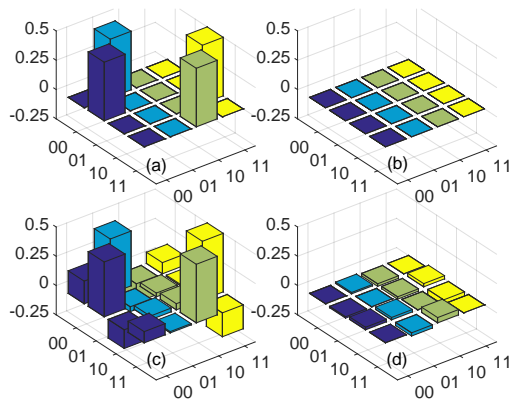}
\caption{Real (a) and imaginary (b) parts of the theoretically expected deviation density matrix for a L\"uders device ($\rho'_{L}$); real (c) and imaginary (d) parts of the experimental deviation density matrix ($\rho'_{\mathrm{exp}}$).}
\label{fig_tomo}
\end{figure}

Fig. \ref{fig_tomo} compares the experimental results with the theoretically expected deviation density matrices. The correlation \cite{Correltn_fidlty_cory}
\begin{eqnarray}
C = \frac{\mathrm{Tr}[\rho'_{L}\rho'_\mathrm{exp}]}{\sqrt{\mathrm{Tr}[\rho_{L}^{\prime 2}]\mathrm{Tr}[\rho_\mathrm{exp}^{\prime 2}]}}
\label{C_corelatn}
\end{eqnarray}
between the theoretical ($\rho'_{L}$,  Eq. \ref{Lud_finalstate}) and
the experimental ($\rho'_\mathrm{exp}$) deviation density matrices was 0.923.  The reduction in the correlation is mainly due to coherent errors caused by imperfect unitary operators, fluctuations in the dipolar coupling constants due to temperature gradients over the sample volume, inhomogeneous RF fields, as well as due to decoherence. 

The correlation expression in Eq. \ref{C_corelatn} is not directly applicable for the null-matrix $\rho'_N$.  Therefore, we replace $\rho'_N$ with random traceless diagonal matrices, and obtained 0.28 as the upper bound for the correlation of $\rho'_\mathrm{exp}$ with $\rho'_N$.
Therefore we conclude that the experimental deviation density matrix is much closer to $\rho'_L$ (Eq. \ref{Lud_finalstate}), and  strongly favors the L\"uders update rule.  

\section{Conclusions}
Quantum measurements, involving probabilistic state collapse and corresponding measurement outcomes, has always been mysterious.  There have been attempts to deduce rules based on phenomenological observations.  According to one of the earliest reduction rules, given by von Neumann, superposition in a degenerate subspace is destroyed by the measurement of the respective degenerate observable.  This rule was later substantially modified by Gerhart L\"uders.  The modified rule, which is most commonly used, implies that superpositions within the degenerate subspaces are preserved under such a measurement.  

A protocol to determine whether a given measuring device is L\"uders or von Neumann was recently formulated by Hegerfeldt and Mayato \cite{Discriminate_von_Luder_protocol}.  In this work, we have adapted this protocol for quantum information systems, and utilize ancilla qubits for performing a desired measurement on system qubits.  Moreover, we describe an NMR experiment, with two system qubits and two ancilla qubits, to discriminate between L\"uders and von Neumann devices.  Within the limitations of experimental NMR techniques, we found that the measurements are of L\"uders type.

There is a possibility that the above measurement is still of von Neumann type, if the chosen initial state happens to be a nondegenerate eigenstate of the actual system observable ($A'$).  One way to rule out this possibility is by changing the initial state (Fig. \ref{HMprotocolfig}).  However, it is also possible that the actual system observable is dynamic, in which case it is even more difficult to discriminate between L\"uders and von Neumann measurements.  In this work we have not excluded these possibilities.
Nevertheless, the present work opens many interesting questions.  For example, how can we build a von Neumann measuring device, or even an intermediate measuring device that partly breaks the degeneracy? More importantly, further research in this direction may throw some light on fundamental aspects of quantum measurement itself.   

\section*{Acknowledgement}
Authors acknowledge useful discussions with Anjusha V. S.  Authors are also grateful to Prof. A. K. Rajagopal and Prof. A. R. Usha Devi for numerous comments and suggestions.
This work was supported by DST/SJF/PSA-03/2012-13 and CSIR-03(1345)/16/EMR-II.

\bibliographystyle{apsrev4-1}
\bibliography{bib_ch}

\begin{thebibliography}{20}
\bibitem [{lud()}]{ludernote1}%
  \BibitemOpen
  \href@noop {} {}\bibinfo {note} {The L\"uders device may also map $A$ to $A'$
  which is of the form $A'=a'_1(\Pi_0+\Pi_1)+a'_2(\Pi_2+\Pi_3)$. Here $A'$ has
  same degenerate subspaces as that of $A$, but arbitrary eigenvalues
  $a'_j$.}\BibitemShut {Stop}%
\end{thebibliography}

\end{document}